\UseRawInputEncoding
\documentclass[letterpaper, 10 pt, conference]{ieeeconf}  
\usepackage{graphicx}      
\usepackage{cite}        
\usepackage{siunitx}
\usepackage{amsmath,cancel}
\usepackage{array}
\usepackage{amsmath}
\usepackage{tikz}
\usepackage{mathdots}
\usepackage{yhmath}
\usepackage{cancel}
\usepackage{color}
\usepackage{multirow}
\usepackage{amssymb}
\usepackage{gensymb}
\usepackage{tabularx}
\usepackage{booktabs}
\usepackage{mathrsfs}
\usepackage{arydshln}
\usepackage{url}

\usepackage[font=footnotesize,labelfont=bf]{caption}
\usepackage[font=footnotesize,labelfont=bf]{subcaption}

\IEEEoverridecommandlockouts                              

\title{\LARGE \bf
Complex-order Reset Control System
}

\author{Nima Karbasizadeh$^{1}$ and S. Hassan HosseinNia$^{1}$
	\thanks{This work was supported by NWO, through OTP TTW project \#16335.}
	\thanks{$^{1}$Department of Precision and Microsystem Engineering, Delft University of Technology, Delft, The Netherlands,
	}
	\thanks{{\tt\small N.KarbasizadehEsfahani@tudelft.nl}}
	\thanks{{\tt\small S.H.HosseinNiaKani@tudelft.nl}}%
}

\begin{document}

\maketitle
\thispagestyle{empty}
\pagestyle{empty}

\begin{abstract}

According to the well-known loop shaping method for the design of controllers, the performance of the controllers in terms of step response, steady-state disturbance rejection and noise attenuation and robustness can be improved by increasing the gain at lower frequencies and decreasing it at higher frequencies and increasing the phase margin as much as possible. However, the inherent properties of linear controllers, the Bode's phase-gain relation, create a limitation. In theory, a complex-order transfer function can break the Bode's gain-phase relation; however, such transfer function cannot be directly implemented and should be approximated. This paper proposes a reset element and a tuning method to approximate a Complex-Order Controller (CLOC) and, through a simulation example, shows the benefits of using such a controller.   

\end{abstract}

\section{INTRODUCTION}
The increasing demands for speed and accuracy from the high-tech industry, especially in the field of precision motion control, has pushed the linear controllers to their inherent limitations, namely - Bode's gain-phase relationship and waterbed effect ~\cite{aastrom2000limitations,schmidt2020design,freudenberg2000survey}. The well-known limitations pose a trade-off between tracking and steady-state precision on one side and bandwidth, stability margins and transient response properties on the other.\\
From the perspective of the loop-shaping technique, the industry-standard method for controller design in the frequency domain, one needs to break this gain-phase relationship to break the trade-off. This was first recognized in complex order derivatives used in the third generation CRONE technique~\cite{oustaloup2000frequency,sabatier2015fractional}. However, such a derivative which can potentially show a negative gain slope with a corresponding positive increasing phase, is unfortunately not practically implementable in the linear domain.  Existing attempts in the literature for approximating such behaviour resulted in unstable poles, non-minimum phase zeros or poor approximation of gain behaviour among other issues~\cite{oustaloup1991commande,moghadam2018tuning}.\\
The impossibility of achieving implementable complex-order behaviour in linear systems has made researchers and industries interested in nonlinear control methods that are industry-compatible in design and implementation. One such interesting concept is reset control, which was first introduced by Clegg~\cite{clegg1958nonlinear}. In~\cite{valerio2019reset,saikumar2019complex}, a method to approximate Complex-Order Controllers (CLOC) using reset control is introduced, which consists of  multiple reset states, which makes the system become unnecessarily complicated and can potentially deteriorate the precision performance of the system.\\
Based on Describing Function (DF) analysis, it is established that the reset integrator proposed by Clegg reduces the phase lag of the integrator by $52^\circ$. Although this already breaks the Bode's gain-phase relation for linear control systems, there are concerns while using Clegg's Integrator (CI) in practice, namely, the accuracy of DF approximation, limit-cycle, etc.~\cite{banos2011reset}. In order to address the drawbacks and exploit the benefits, the idea was later extended to more sophisticated elements such as ``First-Order Reset Element''~\cite{horowitz1975non,krishnan1974synthesis} and ``Second-Order Reset Element''~\cite{hazeleger2016second} or using Clegg's integrator in the form of PI+CI~\cite{banos2007definition} or resetting the state to a fraction of its current value, known as partial resetting~\cite{beker2004fundamental}.\\
One of the recent studies introduces a new reset element called ``Constant-in-Gain, Lead-in-Phase'' (CgLp)~\cite{saikumar2019constant}. DF analysis of this element shows that it can provide broadband phase lead while maintaining a constant gain. Such an element is used in the literature to replace some part of the differentiation action in PID controllers as it will help improve the precision of the system according to the loop-shaping concept~\cite{karbasizadeh2021fractional,karbasizadeh2020benefiting,dastjerdi2021frequency,saikumar2019constant}. Despite the fact that the analysis and designs are focused on describing functions without taking higher-order harmonics into account, significant tracking and steady-state precision improvements are reported.\\
The main contribution of this paper is to propose a reset controller to approximate a complex-order transfer function. This reset controller is based on the CgLp structure with a shaped reset signal. The CgLp which is used has only one resetting state, which reduces the complications  of multiple resetting states. The shaping filter and its tuning method for the reset signal will be introduced to tune the slope of gain and phase as in a complex-order transfer function. This paper also shows the benefits of using such a reset controller in step responses and steady-state precision over a linear PID controller in simulation.\\
The remainder of the paper is organized as follows: Section II introduces the preliminaries of the study. Section III will introduce the reset element to approximate the complex-order transfer function. Section IV will show an illustrative example and, finally, paper will be closed with conclusions and future work tips.
\section{Preliminaries}
This section will discuss the preliminaries of this study.
\subsection{General Reset Controller}
The general form of a SISO reset controllers used in this study is as following:
\begin{align}
	\label{eq:reset}
	{{\sum }_{R}}:=\left\{ \begin{aligned}
		& {{{\dot{x}}}_{r}}(t)={{A_r}}{{x}_{r}}(t)+{{B_r}}e(t),&\text{if }e(t)\ne 0\\ 
		& {{x}_{r}}({{t}^{+}})={{A}_{\rho }}{{x}_{r}}(t),&\text{if }e(t)=0 \\ 
		& u(t)={{C_r}}{{x}_{r}}(t)+{{D_r}}e(t) \\ 
	\end{aligned} \right.
\end{align}
where $A_r,B_r,C_r,D_r$ denote the state space matrices of the Base Linear System (BLS) and reset matrix is denoted by $A_\rho=\text{diag}(\gamma_1,...,\gamma_n)$ which contains the reset coefficients for each state. $e(t)$ and $ u(t) $ represent the input and output for the reset controller, respectively.
\subsection{Describing Functions}
Describing function analysis is the known approach in literature for approximation of frequency response of nonlinear systems like reset controllers~\cite{guo2009frequency}. However, the DF method only takes the first harmonic of Fourier series decomposition of the output into account and neglects the effects of the higher-order harmonics. This simplification can be significantly inaccurate under certain circumstances~\cite{karbasizadeh2020benefiting}. The ``Higher-Order Sinusoidal Input Describing Function'' (HOSIDF) method has been introduced in~\cite{nuij2006higher} to provide more accurate  information about the frequency response of nonlinear systems by investigation of higher-order harmonics of the Fourier series decomposition. In other words, in this method, the nonlinear element will be replaced by a virtual harmonic generator.\\
Per definition, describing functions are calculated for sinusoidal inputs. Thus, assuming $e(t)=\sin(\omega t)$, HOSIDF method was developed in~\cite{saikumar2021loop,dastjerdi2020closed} for reset elements defined by~\eqref{eq:reset} as follows:
\begin{align}  \nonumber
	\label{eq:hosidf}
	& H_n(\omega)=\left\{ \begin{aligned}
		& C_r{{(j\omega I-A_r)}^{-1}}(I+j{{\Theta }}(\omega ))B_r+D_r,~~ n=1\\ 
		& C_r{{(j\omega nI-A_r)}^{-1}}j{{\Theta }}(\omega )B_r,\quad~~~~~~\text{odd }n> 2\\ 
		& 0,\qquad\qquad\qquad\qquad\quad\qquad~\quad~~~~\text{even }n\ge 2\\ 
	\end{aligned} \right. \\
	&\begin{aligned}
		& {{\Theta }}(\omega )=-\frac{2{{\omega }^{2}}}{\pi }\Delta (\omega )[{{\Gamma }}(\omega )-{{\Lambda }^{-1}}(\omega )] \\  
		& \Lambda (\omega )={{\omega }^{2}}I+{{A_r}^{2}} \\  
		& \Delta (\omega )=I+{{e}^{\frac{\pi }{\omega }A_r}} \\  
		& {{\Delta }_{\rho}}(\omega )=I+{{A}_{\rho}}{{e}^{\frac{\pi }{\omega }A_r}} \\  
		& {{\Gamma }}(\omega )={\Delta }_{\rho}^{-1}(\omega ){{A}_{\rho}}\Delta (\omega ){{\Lambda }^{-1}}(\omega ) \\
	\end{aligned} 
\end{align}
where $H_n(\omega)$ is the $n^{\text{th}}$ harmonic describing function for sinusoidal input with the frequency of $\omega$. It has to noted that according to~\cite{guo2009frequency}, the convergence and asymptotic stability of reset elements in open-loop is guaranteed when $\vert \lambda(A_\rho)\vert<1$, where $\lambda(A_\rho)$ stands for eigenvalues of the matrix $A_\rho$.
\subsection{Describing Functions with Shaped Reset Signal}
\begin{figure}[t]
	\vspace{5pt}
	\centering
	\begin{subfigure}{\columnwidth}
		\centering
		\resizebox{0.6\columnwidth}{!}{

		\tikzset{every picture/.style={line width=0.75pt}} 
		
		\begin{tikzpicture}[x=0.75pt,y=0.75pt,yscale=-1,xscale=1]
			
			\draw  [fill={rgb, 255:red, 0; green, 0; blue, 0 }  ,fill opacity=0.05 ][dash pattern={on 4.5pt off 4.5pt}] (165,33.38) .. controls (165,23.23) and (173.23,15) .. (183.38,15) -- (306.63,15) .. controls (316.77,15) and (325,23.23) .. (325,33.38) -- (325,101.63) .. controls (325,111.77) and (316.77,120) .. (306.63,120) -- (183.38,120) .. controls (173.23,120) and (165,111.77) .. (165,101.63) -- cycle ;
			\draw   (239.37,53.48) -- (293.13,53.48) -- (293.13,97.7) -- (239.37,97.7) -- cycle ;
			\draw    (228.5,109.48) -- (301.84,38.57) ;
			\draw [shift={(304,36.48)}, rotate = 495.96] [fill={rgb, 255:red, 0; green, 0; blue, 0 }  ][line width=0.08]  [draw opacity=0] (10.72,-5.15) -- (0,0) -- (10.72,5.15) -- (7.12,0) -- cycle    ;
			
			\draw    (125,75) -- (234.9,75) ;
			\draw [shift={(237.9,75)}, rotate = 180] [fill={rgb, 255:red, 0; green, 0; blue, 0 }  ][line width=0.08]  [draw opacity=0] (8.93,-4.29) -- (0,0) -- (8.93,4.29) -- cycle    ;
			\draw    (294,75) -- (360,75) ;
			\draw  [dash pattern={on 4.5pt off 4.5pt}]  (175,75) -- (175,95) -- (240,95) ;
			
			\draw (144,60) node    {$e( t)$};
			\draw (209,104) node    {$x_{rl}(t)$};
			\draw (312.5,31) node    {$A_{\rho }$};
			\draw (349.5,60) node    {$u( t)$};
			\draw (266.25,72.98) node    [font=\Large]  {$\sum _{R}$};

		\end{tikzpicture}
		}
		\caption{A conventional reset element. Arrow indicates the resetting action. The resetting action is determined by $x_{rl}(t)$ which is equal to $e(t)$, i.e., the resetting condition is $x_{rl}(t)=e(t)=0$.}
		\label{fig:reset_conv}
	\end{subfigure}\\
	\begin{subfigure}{\columnwidth}
		\centering
		\resizebox{0.6\columnwidth}{!}{
	
		\tikzset{every picture/.style={line width=0.75pt}} 
		
		\begin{tikzpicture}[x=0.75pt,y=0.75pt,yscale=-1,xscale=1]
			
			\draw  [fill={rgb, 255:red, 0; green, 0; blue, 0 }  ,fill opacity=0.05 ][dash pattern={on 4.5pt off 4.5pt}] (95,40.38) .. controls (95,26.36) and (106.36,15) .. (120.38,15) -- (299.63,15) .. controls (313.64,15) and (325,26.36) .. (325,40.38) -- (325,134.63) .. controls (325,148.64) and (313.64,160) .. (299.63,160) -- (120.38,160) .. controls (106.36,160) and (95,148.64) .. (95,134.63) -- cycle ;
			\draw   (239.37,53.48) -- (293.13,53.48) -- (293.13,97.7) -- (239.37,97.7) -- cycle ;
			\draw    (228.5,109.48) -- (301.84,38.57) ;
			\draw [shift={(304,36.48)}, rotate = 495.96] [fill={rgb, 255:red, 0; green, 0; blue, 0 }  ][line width=0.08]  [draw opacity=0] (10.72,-5.15) -- (0,0) -- (10.72,5.15) -- (7.12,0) -- cycle    ;
			
			\draw    (60,75) -- (234.9,75) ;
			\draw [shift={(237.9,75)}, rotate = 180] [fill={rgb, 255:red, 0; green, 0; blue, 0 }  ][line width=0.08]  [draw opacity=0] (8.93,-4.29) -- (0,0) -- (8.93,4.29) -- cycle    ;
			\draw    (294,75) -- (360,75) ;
			\draw  [dash pattern={on 4.5pt off 4.5pt}]  (115,75) -- (115,135) -- (133,135) ;
			\draw  [dash pattern={on 4.5pt off 4.5pt}]  (190,135) -- (220,135) -- (220,90) -- (240,90) ;
			\draw   (133.5,114.07) -- (189.49,114.07) -- (189.49,152) -- (133.5,152) -- cycle ;

			\draw (76,60) node    {$e( t)$};
			\draw (204,124) node   [font=\small] {$x_{rl}(t)$};
			\draw (312.5,31) node    {$A_{\rho }$};
			\draw (349.5,60) node    {$u( t)$};
			\draw (162,99.09) node  [font=\footnotesize] [align=left] {\begin{minipage}[lt]{32.67pt}\setlength\topsep{0pt}
					\begin{flushright}
						\textit{Shaping}
					\end{flushright}
					\begin{center}
						\textit{Filter}
					\end{center}
					
			\end{minipage}};
			\draw (162,135) node  [font=\Large]  {$SF( s)$};
			\draw (266.25,72.98) node   [font=\Large] {$\sum _{R}$};

		\end{tikzpicture}
	}
		\caption{A reset element with shaped reset signal. Arrow indicates the resetting action. The resetting action is determined by $x_{rl}(t)$ which is not equal to $e(t)$, i.e., the resetting condition is $x_{rl}(t)=0$.}
		\label{fig:reset_shaped}
	\end{subfigure}\\
	
	\caption{A conventional reset element vs. a reset element with shaped reset signal. }
	\label{fig:reset_elements}
\end{figure}
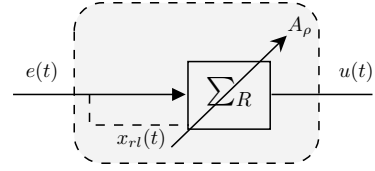
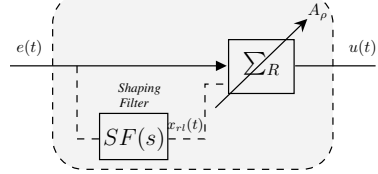

In a conventional reset element, the reset condition is based on the input to the reset element, i.e., $e(t)$. However, one can use a signal other than the input for reset condition. Denoting the reset signal as $x_{rl}(t)$, the reset condition will change to $x_{rl}(t)=0$. This paper creates a reset signal by filtering $e(t)$ and thus named shaped reset signal. Assuming $e(t)=\sin(\omega t)$ for HOSIDF analysis purposes, the reset instants will be $t_k=\frac{k \pi }{\omega}$. However, if one creates $x_{rl}(t)$ such that it has $\varphi$ phase shift compared to $e(t)$, or in other terms
\begin{align}
	\label{eq:reset_phi}
	{{\sum }_{R}}:=\left\{ \begin{aligned}
		& {{{\dot{x}}}_{r}}(t)={{A_r}}{{x}_{r}}(t)+{{B_r}}e(t),&\text{if }\sin(\omega t + \varphi)\ne 0\\ 
		& {{x}_{r}}({{t}^{+}})={{A}_{\rho }}{{x}_{r}}(t),&\text{if }\sin(\omega t + \varphi)=0 \\ 
		& u(t)={{C_r}}{{x}_{r}}(t)+{{D_r}}e(t), \\ 
	\end{aligned} \right.
\end{align}
it means that the reset instants will become $t_k=\frac{k \pi + \varphi}{\omega}$,  while maintaining the input, $e(t)$. In this case, the HOSIDF will change to~{\cite{dastjerdi2020closed}}:
\begin{align}  \nonumber
	\label{eq:hosidf_phi}
	&G_{\varphi n}(\omega)=\left\{ \begin{aligned}
		&C_r{{(A_r-j\omega I)}^{-1}}{{\Theta }_{\varphi}}(\omega )&\\
		&\qquad+C_r(j\omega I-A_r)^{-1}B_r+D_r,~~~~~~ n=1\\ 
		& C_r{{(A_r-j\omega nI)}^{-1}}{{\Theta }_{\varphi}}(\omega ),~~~~~\qquad\text{odd }n> 2\\ 
		& 0\qquad\qquad\qquad\qquad\qquad\qquad~~\quad\text{ even }n\ge 2\\ 
	\end{aligned} \right. \\
	&\begin{aligned}
		&{{\Theta }_{\varphi }}(\omega )=\frac{-2j\omega {{e}^{j\varphi }}}{\pi } \Omega (\omega ) \left( \omega I\cos (\varphi )-A_r\sin (\varphi ) \right){{\Lambda }^{-1}}(\omega )B\\
		&\Omega (\omega )=\Delta (\omega )-\Delta (\omega )\Delta _{\rho}^{-1}(\omega ){{A}_{\rho }}\Delta (\omega ).\\
	\end{aligned}
\end{align}
The above indicates that first and higher-order describing functions can be changed by shaping the reset signal. Fig.~\ref{fig:reset_elements} shows a conventional reset element vs. a reset element with shaped reset signal. It has to be noted that for $x_{rl}(t)$, only zero-crossings matter and thus, ideally, its gain does not have any effect on the system.
\subsection{$H_\beta$ condition}
$H_\beta$ condition~\cite{beker2004fundamental}, despite its conservativeness, has gained attention among different criteria for stability of reset control systems ~\cite{banos2011reset,Guo:2015,nevsic2008stability}, because of simplicity and frequency domain applicability. In~\mbox{~\cite{dastjerdi2021newstability}}, the $H_\beta$ condition has been reformulated such that the frequency response functions of the controllers and the plant can be used directly. This method especially includes the case where the reset element is not the first element in the loop.
\subsection{Frequency Response of $s^{\alpha+j\beta }$}
A derivative of complex order can be defined in a variety of ways~\cite{podlubny1998fractional,samko1993fractional,valerio2013introduction}, but it is commonly indicated by the operator $D^{\alpha+j\beta}$.  It makes no difference which one is utilized for the purpose of this paper, as long as the Laplace transform of a derivative of order $\alpha+j \beta \in \mathbb{C}$ generates a complex power of the Laplace variable $s$:
\begin{equation}
	\mathscr{L}\left[D^{\alpha+j \beta} f(t)\right]=s^{\alpha+j \beta} \mathscr{L}[f(t)]
\end{equation}
The initial conditions for Laplace are also irrelevant for this paper, since the frequency response of the the simplest corresponding complex-order transfer function is under consideration, i.e., $G(s)=s^{\alpha+j\beta}$. The frequency response is given by~\cite{valerio2019reset}:
\begin{align}
\label{eq:complex}
	&\begin{aligned}
		G(j \omega) &=j^{\alpha} \omega^{\alpha} j^{j \beta} \omega^{j \beta} \\
		&=\left(\cos \frac{\alpha \pi}{2}+j \sin \frac{\alpha \pi}{2}\right) \omega^{\alpha} \\
		& \times e^{-\frac{\beta \pi}{2}}(\cos (\beta \log \omega)\\
		&+j \sin (\beta \log \omega)) 
	\end{aligned}\\
\label{eq:complex_gain}
	&\begin{aligned}
		20 \log _{10}|G(j \omega)|&=20 \log _{10}\left(\omega^{\alpha} e^{-\frac{\beta \pi}{2}}\right) \\
		&=20 \alpha \log _{10} \omega+20 \log _{10} e^{-\frac{\beta \pi}{2}} 
	\end{aligned}\\
\label{eq:complex_phase}
	&\begin{aligned}
		\arg G(j \omega)&=\angle\left\{\left(\cos \frac{\alpha \pi}{2}+j \sin \frac{\alpha \pi}{2}\right)\right.\\
		&\left.\times[\cos (\beta \log \omega)+j \sin (\beta \log \omega)]\right\} \\
		&=\frac{\alpha \pi}{2}+\beta \log (10) \log _{10} \omega
	\end{aligned}
\end{align}
When $\alpha < 0$ and $\beta >0$, the frequency response will show a negative gain slope and a positive phase slope, for which there is no practical implementation method in linear domain. However, such frequency response is highly desirable especially in precision motion control since one can for example increase the bandwidth of the system without sacrificing the phase margin~\cite{saikumar2019complex}.   
\section{Approximating the Complex-order Behaviour using CgLp}
\begin{figure}[t]
	\centering
	\includegraphics[width=\columnwidth]{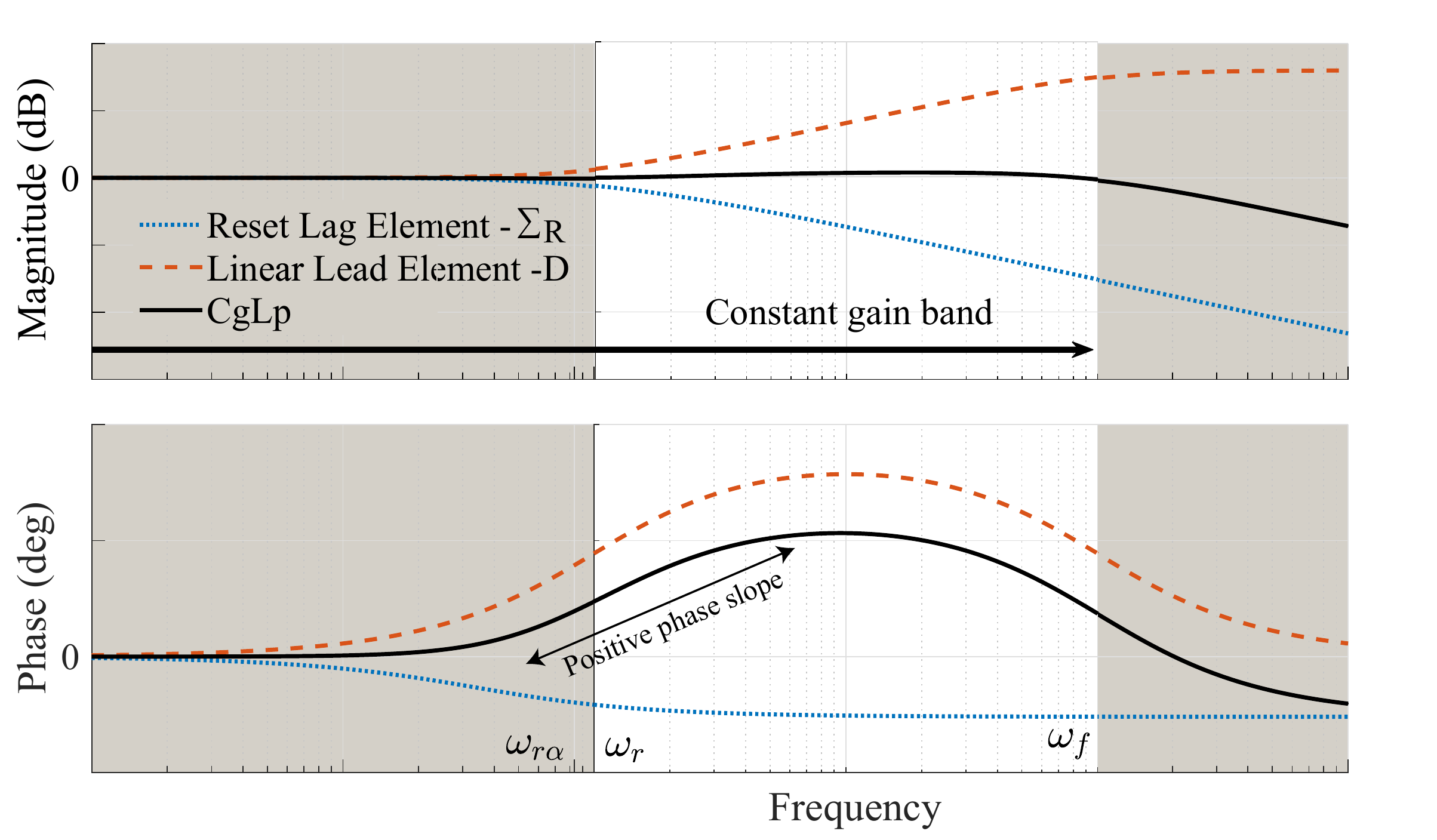}
	\caption{The concept of using combination of a reset lag and a linear lead element to form a CgLp element. The figure is  from~\cite{saikumar2019constant}.}
	\label{fig:cglp}
\end{figure}

$G(s)=s^{\alpha+j\beta}$ is a complex-order transfer function that may be written as $G(s)=s^{\alpha}s^{j\beta}$. A large literature exists on approximation $s^\alpha$ for any non-integer $\alpha \in \mathbb{R}$, with CRONE approximation being one of the more well-known methods. The novelty, on the other hand, resides in the approximation of  $s^{j\beta}$. According to~\eqref{eq:complex} to~\eqref{eq:complex_phase}, for $\beta>0$, such transfer function should show a constant gain behaviour while having a positive phase slope of $\beta \log (10)$.\\
CgLp is a reset element that shows a DF behaviour similar to required. This element creates a unity gain with phase lead in a desired range of frequencies. CgLp can be created using a reset first-order lag filter ${\sum}_R$, a.k.a. FORE,  in series with a corresponding linear first-order lead filter $D(s)$ having the same cut-off frequency as given below.
\begin{align}
	\label{eq:fore}
	&{\sum}_R=\cancelto{\gamma}{\frac{1}{{s}/{{{\omega }_{r }}+1}\;}},&D(s)=\frac{{s}/{{{\omega }_{r\kappa}}}\;+1}{{s}/{{{\omega }_{f}}}\;+1}
\end{align}
where $\omega_{r\kappa}=\kappa \omega_r$, $\kappa$ is a tuning parameter accounting for a shift in corner frequency of the filter due to resetting action,  and $[\omega_{r},\omega_{f}]$ is the frequency range where the CgLp will provide the required phase lead~\cite{saikumar2019constant}. The arrow indicates the resetting action as described in~\eqref{eq:reset}, i.e., the element's state is multiplied by $\gamma$ when the reset condition is met. \\
While the gain of the reset lag element and the linear lead cancel each other to create a unity gain, a phase lead is created due to reduced phase lag in the reset lag filter compared to its linear counterpart. The concept is presented in~Fig.~\ref{fig:cglp}.\\
Fig.~\ref{fig:cglp} shows that within a range of frequency, CgLp shows a unity gain and a positive phase slope resembling the frequency response of $s^{j\beta}$. Although one can tune the phase slope by tuning $\gamma$ (since decreasing $\gamma$ increases the created phase lead), the range of achievable slopes and the range of frequencies for which the slope is constant are limited.\\
An approach to gain more freedom in shaping the positive phase slope in the CgLp element is to shape the phase of the reset element without affecting the gain. In~\cite{karbasizadeh2020bandpassing} a method called ``Phase Shaping'' is presented to shape the phase of a reset controller. The method is based on the fact that according to~\eqref{eq:hosidf_phi}, one can shape the first and higher-order describing function of a reset element by changing $\varphi$. In~\cite{karbasizadeh2020bandpassing}, the objective of changing $\varphi$ is achieved through putting a shaping filter on the input signal of the reset element while maintaining the reset signal the same.\\
However, in this paper, the objective is achieved by using the shaping filter on the reset signal itself, as depicted in Fig.~\ref{fig:reset_shaped}. In this configuration, one can see that 
\begin{equation}
	\label{eq:varphi_sf}
	\varphi=\angle SF(j\omega).
\end{equation}
If $\sum_{R}$ is a FORE as presented in~\eqref{eq:fore}, and by defining
\begin{equation}
	\label{eq:psi}
	\psi:=-\varphi-\tan^{-1}\left( \frac{\omega}{\omega_r} \right), 
\end{equation} 
one can conclude the following points, following the same logic and procedure as presented in~\cite{karbasizadeh2020bandpassing}:
\begin{itemize}
	\item If 
	\begin{equation}
		\varphi =-{{\tan }^{-1}}\left( \frac{\omega }{{{\omega }_{r}}} \right)\Rightarrow \psi =0,
	\end{equation}
 	then in the steady-state, the resetting action will not affect the output since the resetting action will happen when the output of the base linear system (BLS) is already zero. Thus, the $x(t)=x(t^+)=0$ at the time of reset. In this case, all the higher-order harmonics will be zero because the system is completely acting linear in steady-state, and obviously, the phase advantage from the resetting action disappears. Furthermore, if one uses such an element to form a CgLp, the element's phase will remain at zero, as the linear lead will cancel out both gain and phase of the reset element.   
	\item As $\psi \to -90^\circ$, the resetting action will create its benefit, and the reset element will provide more phase advantage. 
 	\item The phase advantage created by the reset element is dependent on $\omega_r$, $\gamma$ and $\psi$, and for frequencies  larger enough than $\omega_r$, i.e., $\omega>10\omega_r$, it only depends on $\gamma$ and $\psi$.
\end{itemize}
According to the points mentioned above, for each value of $\gamma \in [-1, 1]$, one can shape the reset element's phase advantage and, thus, the CgLp's phase by shaping $\psi$. In other words, by designing the shaping filter, $SF(s)$, one can achieve the desired phase slope for the CgLp.\\
\section{Design of the Shaping Filter}
\label{sec:desgin}
Let
\begin{equation}
	SF(s)=Q(s)K(s) \quad \text{and} \quad K(s)=\frac{1}{s/\omega_r+1}.
\end{equation}
Thus according to~\eqref{eq:varphi_sf} and~\eqref{eq:psi},
\begin{equation}
	\psi=-\angle SF(s)-\tan^{-1}\left( \frac{\omega }{{{\omega }_{r}}} \right)=-\angle Q(s).
\end{equation}
Now shaping $\psi$ and thus the phase advantage of CgLp has been reduced to shaping the $\angle Q(s)$.\\
The requirements for $SF(s)$ can be categorized into three regions:
\begin{itemize}
	\item At the lower frequency region, which is critical for sinusoidal tracking and disturbance rejection performance, the higher-order harmonics should be reduced as much as possible~\cite{karbasizadeh2020benefiting,karbasizadeh2021fractional,karbasizadeh2020bandpassing,bahnamiri2020tuning}. Meaning that $\psi$ and thus $\angle Q(s)$ should remain as close as possible to $0^\circ$.
	\item At cross-over frequency region that will be presented as $[\omega_l,\omega_h]$, $\angle Q(s)$ should increase to $90^\circ$ for CgLp to create phase advantage. Moreover, the increase slope should be tuned for CgLp to maintain a positive phase slope of $\beta \log\left(10\right)$.
	\item At higher frequency region, the gain of $SF(s)$ should have a negative slope to attenuate the high-frequency content of reset signal and thus avoid excessive reset action due to noise. It has already been taken care of by the presence of $K(s)$.
\end{itemize} 
In order to achieve a $Q(s)$ with tunable phase slope, one may refer to~\cite{valerio2013variable}, where it introduces a variation of CRONE approximation as follows
\begin{equation}
	\label{eq:Q}
	\begin{aligned}
		&Q(s)= \frac{\prod_{n=1}^{M}\left(1+\frac{s}{\omega_{z n}}\right)}{\prod_{n=1}^{N}\left(1+\frac{s}{\omega_{p n}}\right)} \\
		&\omega_{z j+1}=\zeta \omega_{z j}, \quad j=1,2, \ldots, M-1 \\
		&\omega_{p i+1}=\eta \omega_{p i}, \quad i=1,2, \ldots, N-1
	\end{aligned}
\end{equation}
where the slope of phase within the range of approximation is 
\begin{equation}
	\label{eq:calG}
	\mathfrak{S}=\frac{\pi}{2} \frac{1}{\log _{10} \zeta}-\frac{\pi}{2} \frac{1}{\log _{10} \eta}~\mathrm{rad} / \text {decade }
\end{equation}
Although $\mathfrak{S}$ is directly related to $\beta$, a numerical optimization problem should be solved to find $\eta$ and $\zeta$ for a desired $\beta$. For approximation to happen in the range of $[\omega_l,\omega_h]$, one should choose $\omega_{z1}=\omega_{p1}=\omega_l$ and choose $N$ and $M$ accordingly to cover the whole range of frequencies. 
\begin{figure}[t]
	\vspace{5pt}
	\centering
	\begin{subfigure}{\columnwidth}
		\includegraphics[width=\columnwidth]{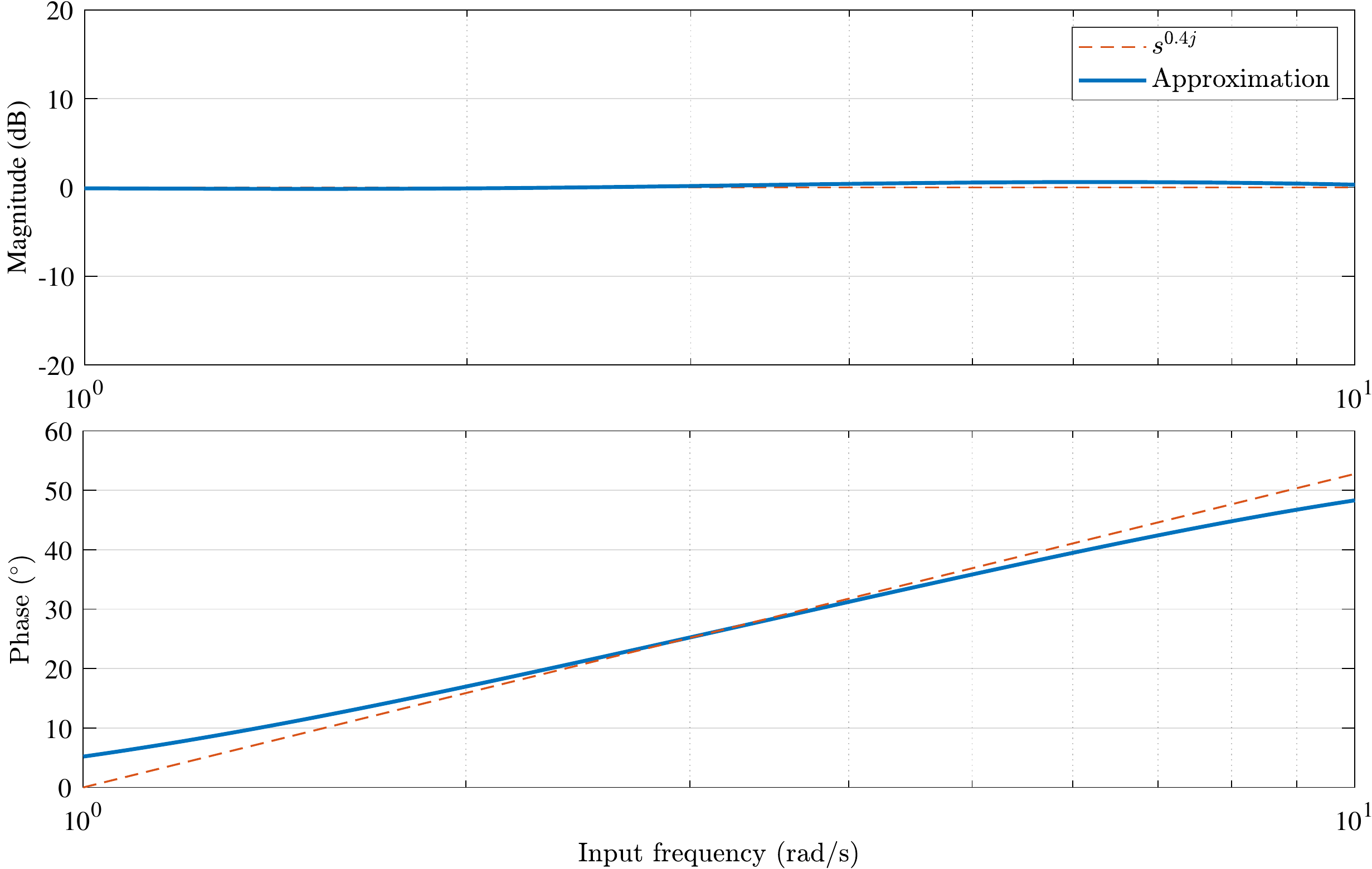}
		\caption{Frequency response of $s^{0.4j}$ and its DF approximation using CgLp in the range of $[1,10]$ rad/s. $N=M=4$, $\zeta=3.72$ and $\eta=21.37$.}
		\label{fig:s0.4j}
	\end{subfigure}\\
	\begin{subfigure}{\columnwidth}
		\centering
		\includegraphics[width=\columnwidth]{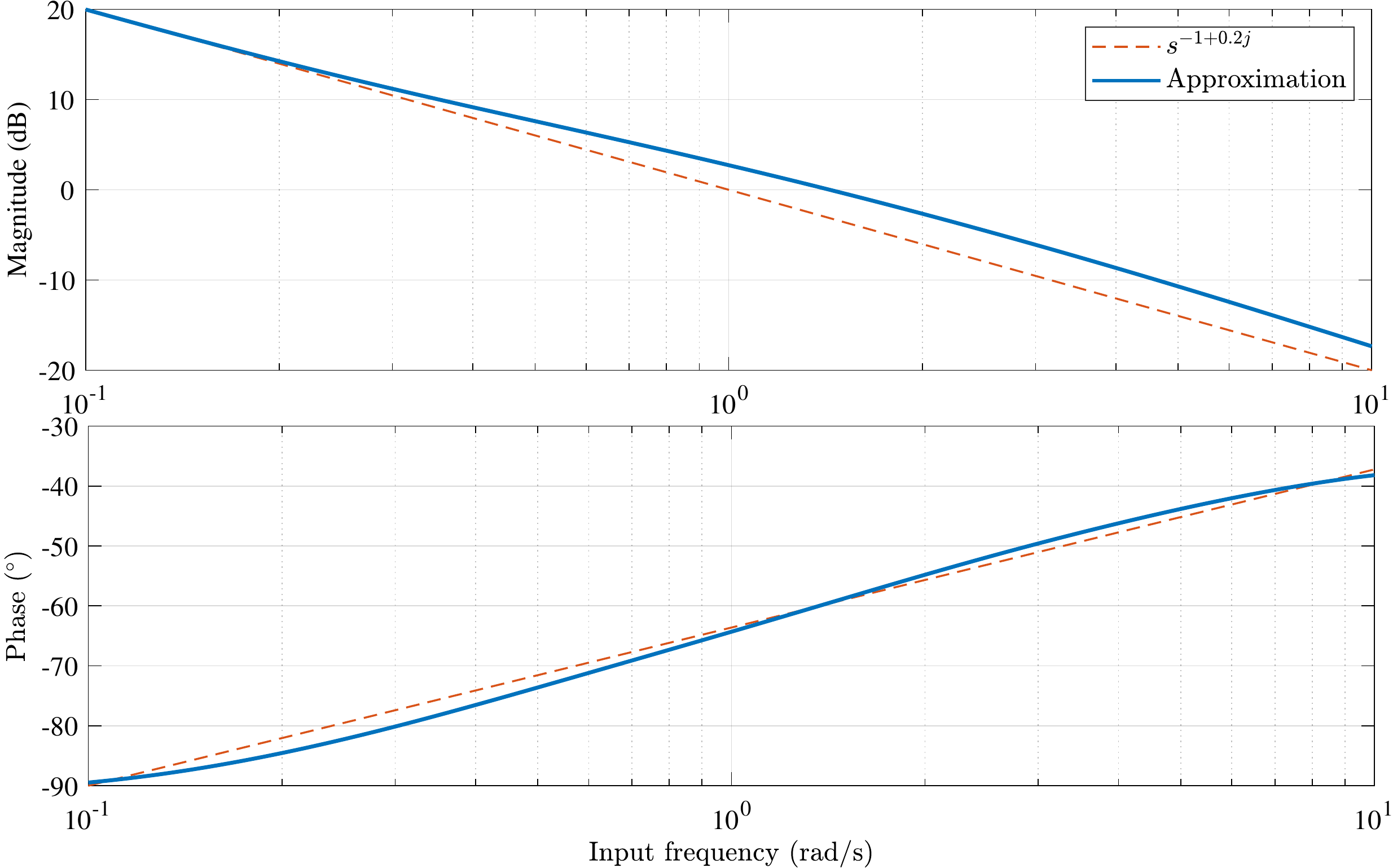}
		\caption{Frequency response of $s^{-1+0.2j}$ and its DF approximation using CgLp in the range of $[0.1,10]$ rad/s. $N=M=7$, $\zeta=2.93$ and $\eta=4.30$.}
		\label{fig:s-1_0.2j}
	\end{subfigure}\\
	
	\caption{A conventional reset element vs. a reset element with shaped reset signal. }
	\label{fig:compolex_example}
\end{figure}
\section{Complex-Order Tamed Differentiation}
\begin{figure}[!t]
	\centering
	\begin{subfigure}{\columnwidth}
		\resizebox{\columnwidth}{!}{

			\tikzset{every picture/.style={line width=0.75pt}} 
			
			\begin{tikzpicture}[x=0.75pt,y=0.75pt,yscale=-1,xscale=1]
				
				\draw  [line width=0.75]  (117.3,50.25) -- (233.3,50.25) -- (233.3,146) -- (117.3,146) -- cycle ;
				\draw  [line width=0.75]  (272.5,63) -- (373.5,63) -- (373.5,131) -- (272.5,131) -- cycle ;
				
				\draw [line width=0.75]    (66.42,96.67) -- (116.22,96.45) ;
				\draw [line width=0.75]    (550,97) -- (550,183) -- (55.5,182.67) -- (55.46,110.73) ;
				\draw [shift={(55.46,107.73)}, rotate = 449.97] [fill={rgb, 255:red, 0; green, 0; blue, 0 }  ][line width=0.08]  [draw opacity=0] (8.93,-4.29) -- (0,0) -- (8.93,4.29) -- cycle    ;
				\draw [line width=0.75]    (-1,96.45) -- (41.5,96.65) ;
				\draw [shift={(44.5,96.67)}, rotate = 180.27] [fill={rgb, 255:red, 0; green, 0; blue, 0 }  ][line width=0.08]  [draw opacity=0] (8.93,-4.29) -- (0,0) -- (8.93,4.29) -- cycle    ;
				\draw [line width=0.75]    (469.5,97) -- (572.5,97) ;
				\draw [shift={(575.5,97)}, rotate = 180] [fill={rgb, 255:red, 0; green, 0; blue, 0 }  ][line width=0.08]  [draw opacity=0] (8.93,-4.29) -- (0,0) -- (8.93,4.29) -- cycle    ;
				\draw [line width=0.75]    (234.5,97) -- (272.5,97) ;
				\draw  [line width=0.75]  (418.5,76.17) -- (470.5,76.17) -- (470.5,121.67) -- (418.5,121.67) -- cycle ;
				\draw [line width=0.75]    (373.5,96.67) -- (418.5,96.67) ;
				\draw   (44.5,96.67) .. controls (44.5,90.56) and (49.41,85.61) .. (55.46,85.61) .. controls (61.52,85.61) and (66.42,90.56) .. (66.42,96.67) .. controls (66.42,102.77) and (61.52,107.73) .. (55.46,107.73) .. controls (49.41,107.73) and (44.5,102.77) .. (44.5,96.67) -- cycle ; \draw   (47.71,88.85) -- (63.21,104.49) ; \draw   (63.21,88.85) -- (47.71,104.49) ;
				
				\draw (17.84,81.76) node  [font=\Large]  {$r( t)$};
				\draw (547.62,79.76) node  [font=\Large]  {$y( t)$};
				\draw (81.35,80.76) node  [font=\Large]  {$e( t)$};
				\draw (445.57,97.58) node  [font=\LARGE]  {$\frac{1}{s^{2}}$};
				\draw (446,58) node  [font=\normalsize] [align=left] {{\fontfamily{ptm}\selectfont {\large Plant}}};
				\draw (41,85) node    {$+$};
				\draw (68,111) node    {$-$};
				\draw (323,97) node  [font=\Large]  {$k_{p}\left( 1+\dfrac{\omega _{i}}{s}\right)$};
				\draw (321,40) node  [font=\normalsize] [align=left] {{\fontfamily{ptm}\selectfont {\large Integrator}}};
				\draw (175,98.13) node  [font=\Large]  {$\left(\dfrac{\frac{s}{\omega _{d}} +1}{\frac{s}{\omega _{t}} +1}\right) s^{j\beta }$};
				\draw (171,25) node  [font=\large] [align=center] {{\fontfamily{ptm}\selectfont Tamed}\\{\fontfamily{ptm}\selectfont Differenatiator}};

			\end{tikzpicture}
	}
		\caption{Control loop for position control of a mass system. For PID, $\beta=0$ and for CLOC $\beta=0.3$.}
		\label{fig:block-lopp}
	\end{subfigure}\\
	\begin{subfigure}{\columnwidth}
		\centering
		\resizebox{\columnwidth}{!}{

			\tikzset{every picture/.style={line width=0.75pt}} 
			
			\begin{tikzpicture}[x=0.75pt,y=0.75pt,yscale=-1,xscale=1]
				
				\draw  [fill={rgb, 255:red, 0; green, 0; blue, 0 }  ,fill opacity=0.05 ][dash pattern={on 4.5pt off 4.5pt}] (110,43) .. controls (110,27.54) and (122.54,15) .. (138,15) -- (489,15) .. controls (504.46,15) and (517,27.54) .. (517,43) -- (517,147) .. controls (517,162.46) and (504.46,175) .. (489,175) -- (138,175) .. controls (122.54,175) and (110,162.46) .. (110,147) -- cycle ;
				\draw   (351.37,53.48) -- (405.13,53.48) -- (405.13,97.7) -- (351.37,97.7) -- cycle ;
				\draw    (340.5,109.48) -- (413.84,38.57) ;
				\draw [shift={(416,36.48)}, rotate = 495.96] [fill={rgb, 255:red, 0; green, 0; blue, 0 }  ][line width=0.08]  [draw opacity=0] (10.72,-5.15) -- (0,0) -- (10.72,5.15) -- (7.12,0) -- cycle    ;
				
				\draw    (200,75) -- (349.9,75) ;
				\draw    (502,75) -- (561,75) ;
				\draw  [dash pattern={on 4.5pt off 4.5pt}]  (226,75) -- (226,135) -- (244,135) ;
				\draw  [dash pattern={on 4.5pt off 4.5pt}]  (302,135) -- (332,135) -- (332,90) -- (352,90) ;
				\draw   (245.5,114.07) -- (301.49,114.07) -- (301.49,160) -- (245.5,160) -- cycle ;
				\draw   (448.37,54.48) -- (502.13,54.48) -- (502.13,98.7) -- (448.37,98.7) -- cycle ;
				\draw    (404,75) -- (449,75) ;
				\draw  [line width=0.75]  (135,35) -- (200,35) -- (200,115) -- (135,115) -- cycle ;
				\draw    (75,75) -- (135,75) ;
				
				\draw (86,60) node  [font=\Large]  {$e( t)$};
				\draw (425,30) node  [font=\LARGE]   {$\gamma $};
				\draw (547.5,60) node  [font=\Large]  {$u( t)$};
				\draw (378.25,72.98) node   [font=\LARGE]  {$\frac{1}{s/\omega _{r} +1}$};
				\draw (274,137.31) node  [font=\LARGE]  {$\frac{Q( s)}{s/\omega _{r} +1}$};
				\draw (274,95) node  [font=\normalsize] [align=left] {\begin{minipage}[lt]{32.67pt}\setlength\topsep{0pt}
						\begin{flushright}
							\textit{Shaping}
						\end{flushright}
						\begin{center}
							\textit{Filter}
						\end{center}
						
				\end{minipage}};
				\draw (475.25,73.98) node [font=\Large]   {$\frac{s/\omega _{r\kappa } +1}{s/\omega _{f} +1}$};
				\draw (167.5,75) node  [font=\LARGE]  {$\dfrac{\frac{s}{\omega _{d}} +1}{\frac{s}{\omega _{t}} +1}$};

			\end{tikzpicture}
		}
		\caption{Block diagram for approximating $\left(\dfrac{\frac{s}{\omega _{d}} +1}{\frac{s}{\omega _{t}} +1}\right) s^{j\beta }$. Approximation will be done in the range of $[\omega_l,\omega_h].$}
		\label{fig:block_tamed}
	\end{subfigure}\\
	
	\caption{Block diagram for closed-loop control of a mass system using integer and complex-order tamed differentiator and approximation of complex-order tamed differentiator using reset control.}
	\label{fig:block}
\end{figure}
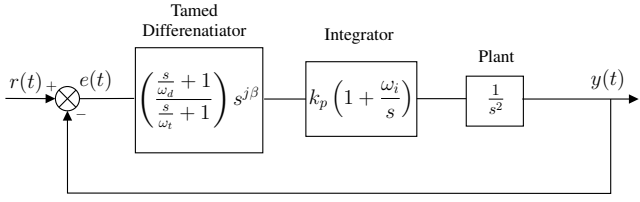
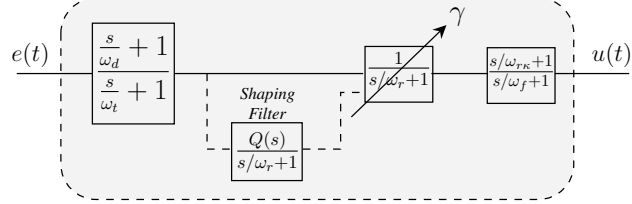
\begin{figure}[t!]
	\centering
	\includegraphics[width=\columnwidth]{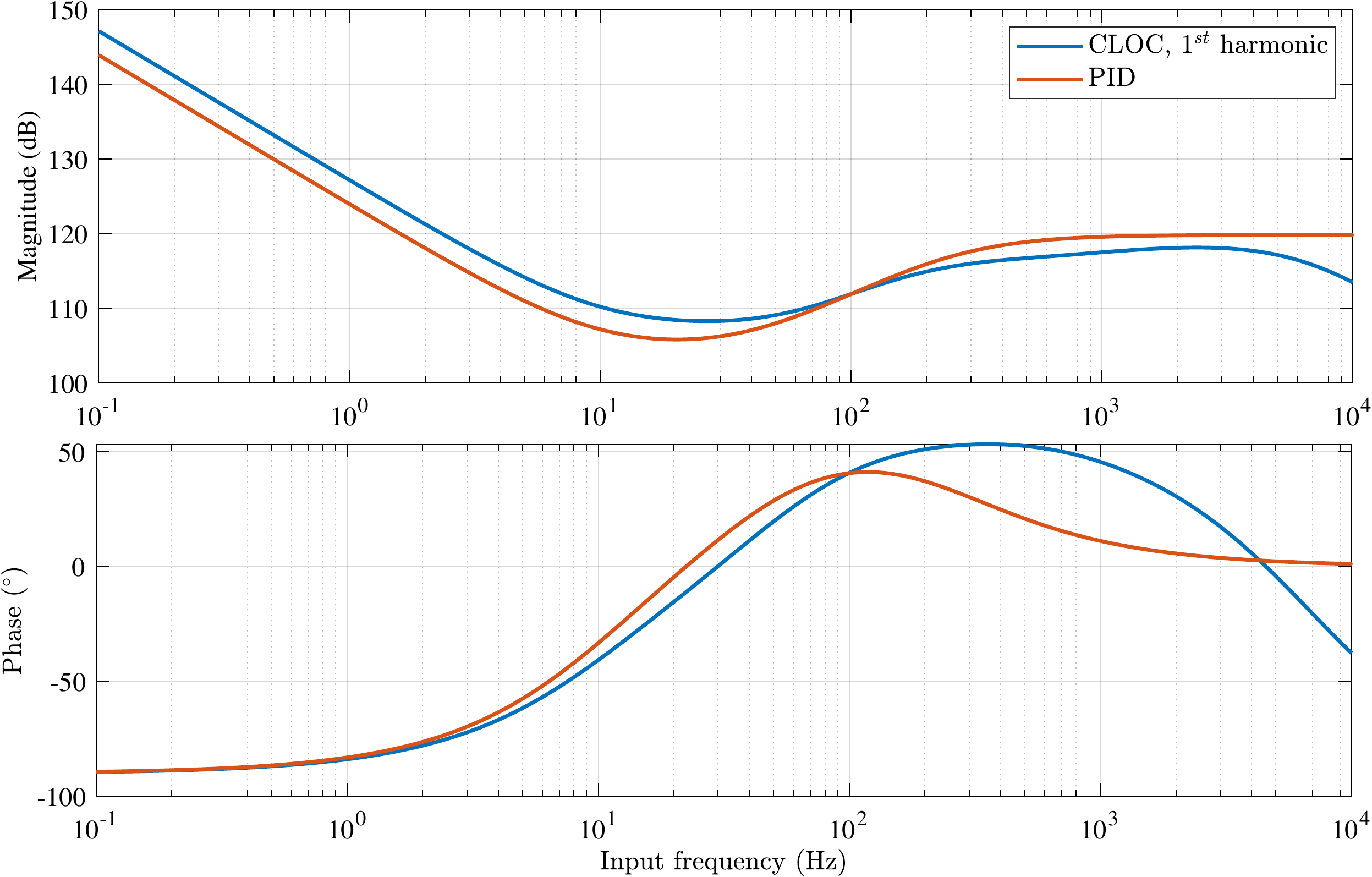}
	\caption{The Bode diagram of PID and the DF of CLOC.}
	\label{fig:controllers}
\end{figure}
One possible application of a complex-order function is to use a complex-order tamed differentiation element instead of linear tamed differentiator~\cite{schmidt2020design} in motion control. For this purpose, without loss of generality, consider a mass system controlled using a control loop presented in Fig.~\ref{fig:block}. Following the steps below, one can design a CLOC 
\begin{enumerate}
	\item Choose $\omega_c$.
	\item Set $\omega_d=\omega_c/1.5$, $\omega_t=1.5\omega_c$ and $\omega_i=\omega_c/10$.
	\item Choose the range of frequencies where the positive slope phase should be present and set $\omega_l$ and $\omega_h$ accordingly.
	\item Choose $\beta$.
	\item Find $\zeta$ and $\eta$ using~\eqref{eq:hosidf_phi},~\eqref{eq:complex_phase} and~\eqref{eq:calG} through optimization. 
	\item Choose $M$ and $N$ s.t. according to~\eqref{eq:Q}, $\omega_{zM}\ge\omega_h$ and  $\omega_{pN}\ge\omega_h$.
	\item Set $\omega_r=\omega_l$ and $\omega_f=10\omega_{h}$.
	\item Choose $\gamma \in (-1,1)$ s.t. the required phase margin is achieved. If not achievable, go to step 4 and correct $\beta$ accordingly. 
\end{enumerate} 
Step 2 ensures that the BLS is stable as a necessary condition for $H_\beta$. It has to be noted that according to waterbed effect, increasing the band of linear differentiation will reduce the steady-state precision of the system. It can be concluded from waterbed effect. Therefore, the band should be minimum only to stabilize the BLS.

As an illustrative example, two controllers will be compared, one with integer-order tamed differentiator, i.e., PID and one with complex-order tamed differentiator, i.e., CLOC.\\
For cross-over frequency, $\omega_c=100$ Hz is chosen, and following the rule of thumb presented in~\cite{schmidt2020design} for tuning PID,
\begin{equation}
	\omega_i=\omega_c/10, \quad \omega_d=\omega_c/2.5, \quad \omega_t=2.5\omega_c, \quad \beta=0.
\end{equation}
Following the steps for designing a CLOC, the following parameters has been tuned for CLOC
 \begin{equation}
 	\omega_i=\omega_c/10, \quad \omega_d=\omega_c/1.5, \quad \omega_t=1.5\omega_c, \quad \beta=0.3.
 \end{equation}
For the approximation of complex-order tamed differentiator, following parameters have been tuned
\begin{align}
	\nonumber
	&\omega_l=0.1^{0.5}\omega_c, \quad \omega_h=10^{0.5}\omega_c, \quad \omega_r=\omega_l, \quad \omega_f=10\omega_h,\\ 
	&N = M = 8, \quad  \zeta=3.314, \quad \eta=7.714, \quad \gamma=0.
\end{align}
\begin{figure}[t]
	\vspace{5pt}
	\centering
	\includegraphics[width=\columnwidth]{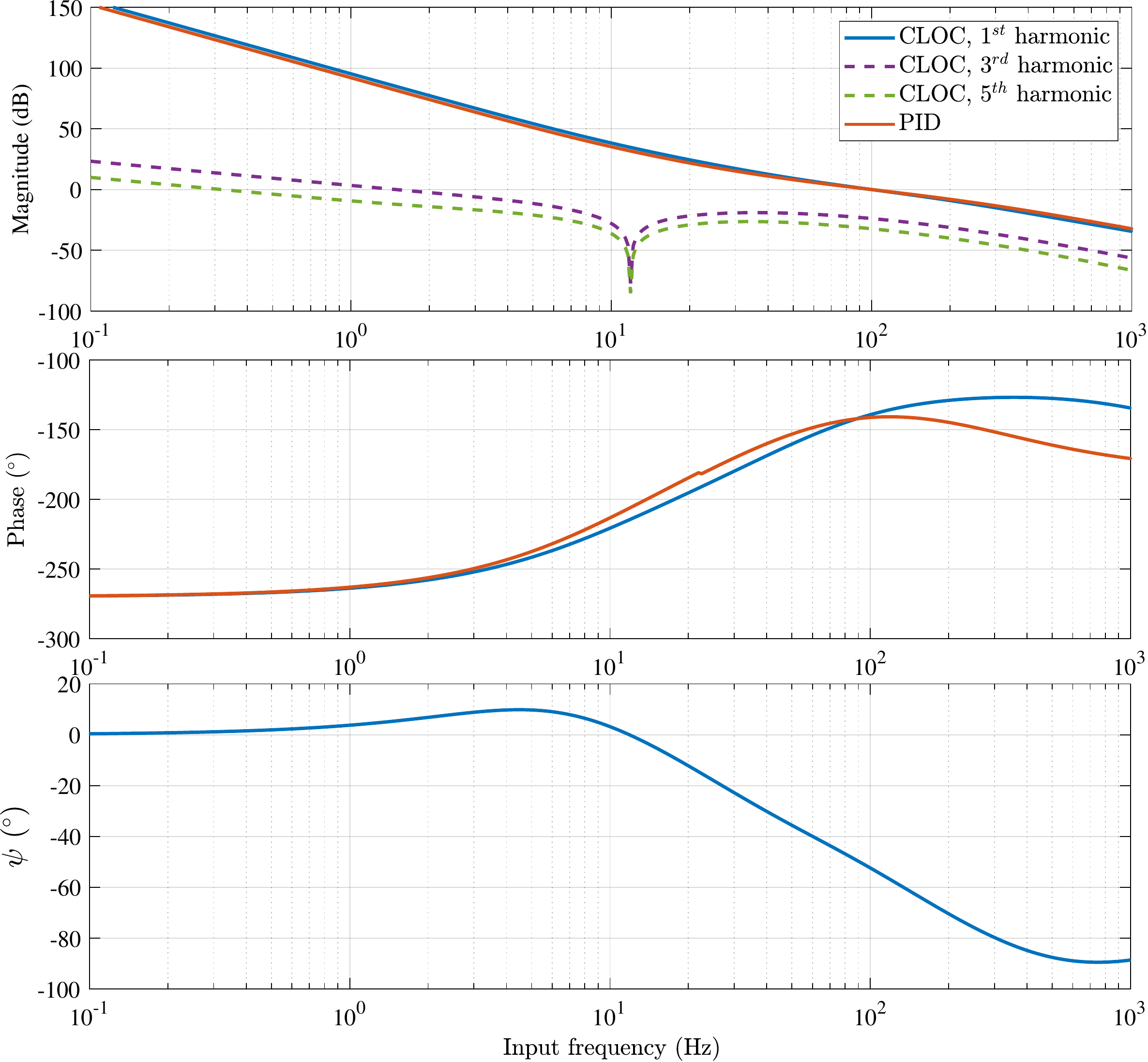}
	\caption{Open-loop Bode diagram of PID and HOSIDF of CLOC including plant.}
	\label{fig:openloop}
\end{figure}
\begin{figure}[t!]
	\centering
	\includegraphics[width=\columnwidth]{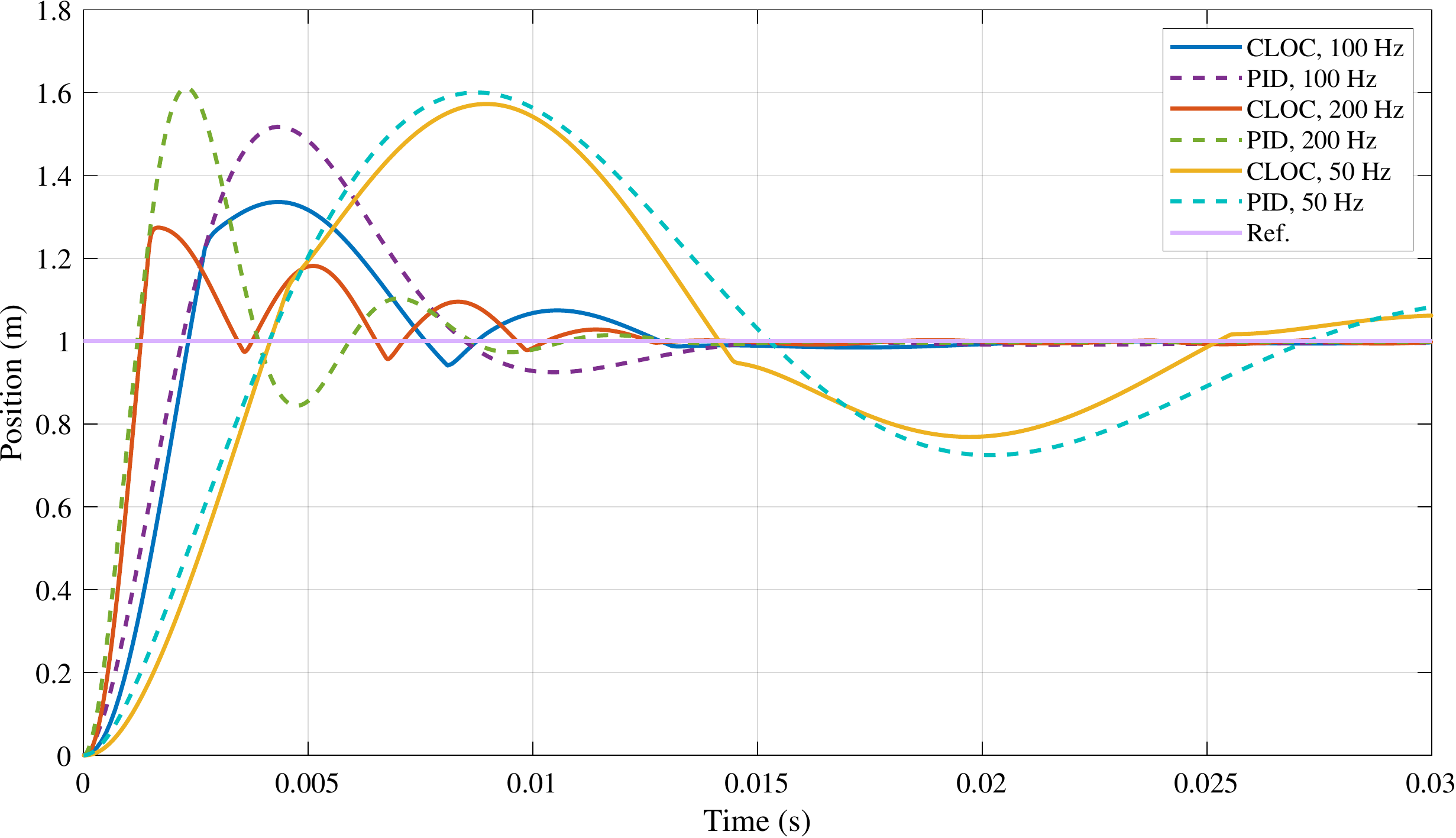}
	\caption{Step response of the CLOC and PID for bandwidth of 50~Hz, 100~Hz and 200~Hz.}
	\label{fig:step}
\end{figure}  
\begin{figure}[t!]
	\centering
	\includegraphics[width=\columnwidth]{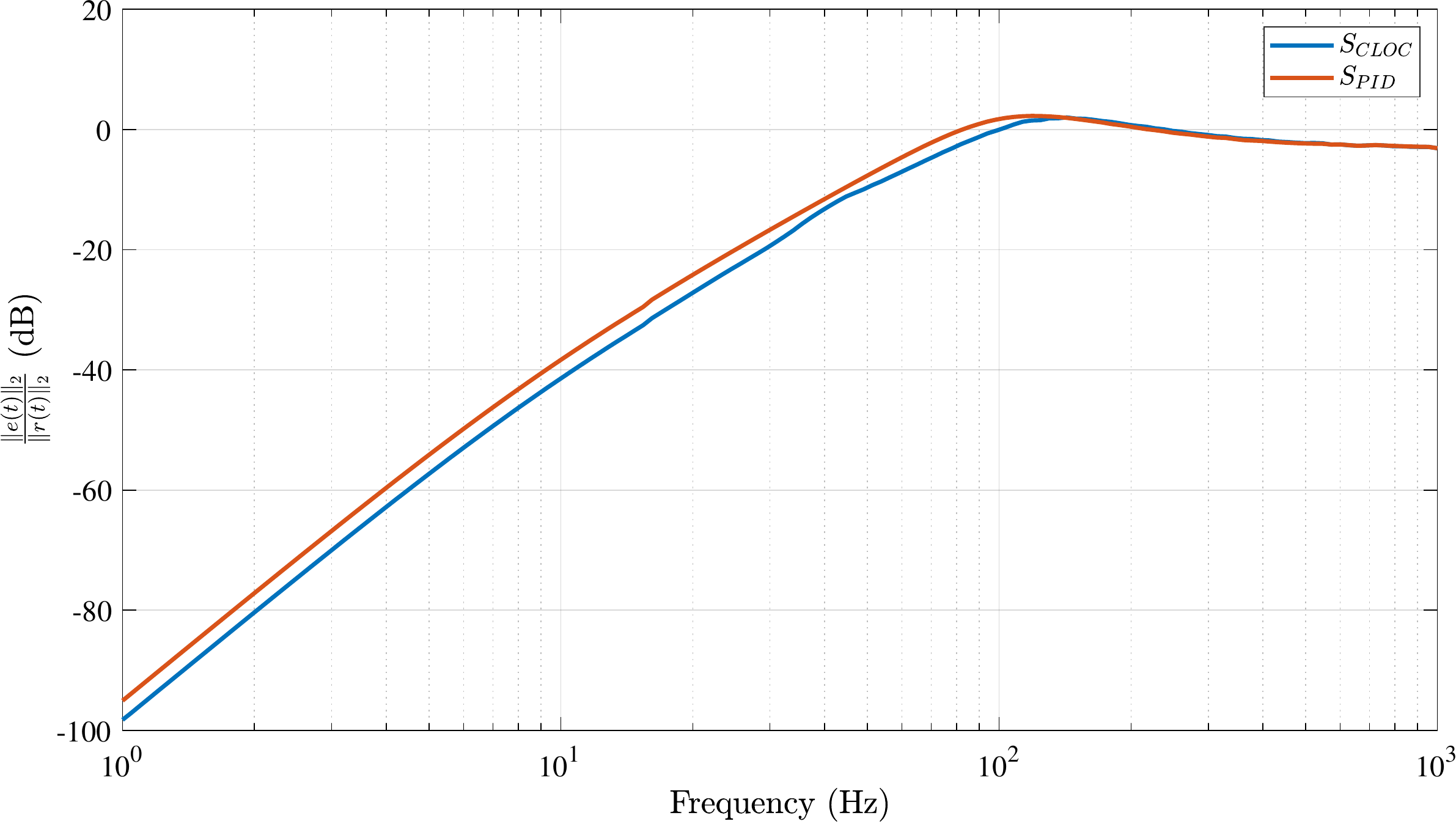}
	\caption{$\frac{\Vert e(t) \Vert_2}{\Vert r(t) \Vert_2}$ plotted for both controllers.}
	\label{fig:sensitivity}
\end{figure}
\begin{figure}[t!]
	\centering
	\includegraphics[width=\columnwidth]{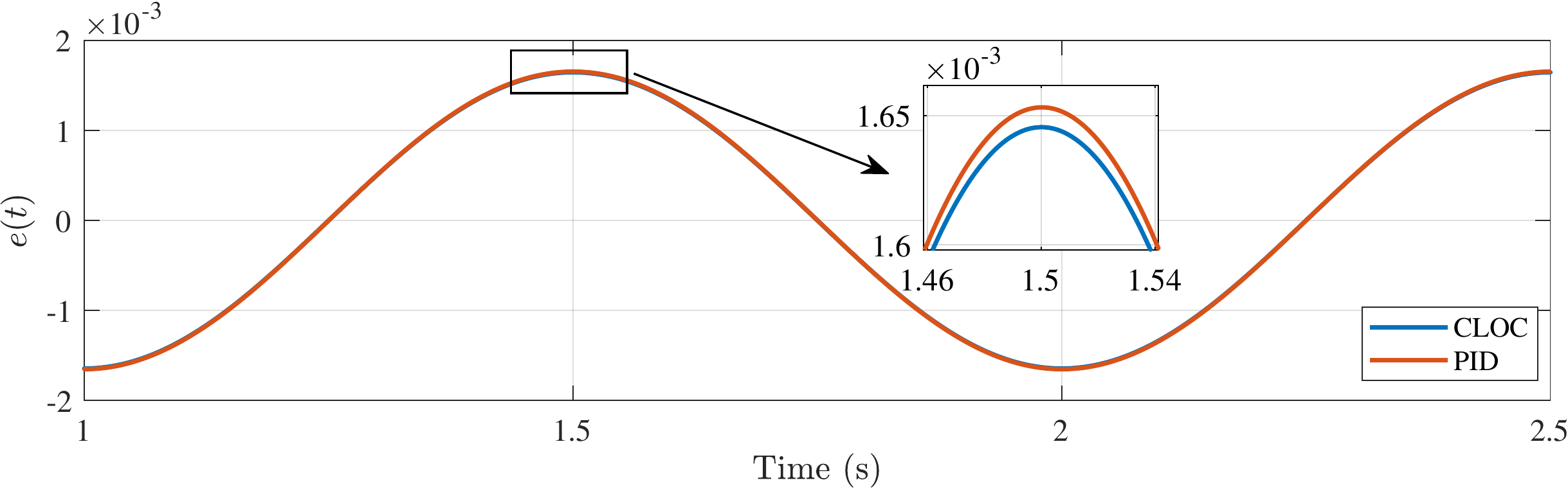}
	\caption{Sinusoidal tracking error of CLOC and PID for $r(t)=\sin(2\pi t)$.}
	\label{fig:sin}
\end{figure}  
TThe above parameters ensure that the positive slope of the phase happens from half a decade before the cross-over frequency to half a decade after.\\
Fig.~\ref{fig:controllers} shows the Bode diagram of the PID controller without plant and DF of CLOC. The figure clearly shows that using a complex-order tamed differentiator in CLOC, the same phase margin as PID is achieved with a lower positive gain slope around the omega, which resulted in higher gain at frequencies below bandwidth and lower gain at frequencies higher than bandwidth compared to PID. Thus, a better steady-state sinusoidal tracking and disturbance rejection and noise attenuation are expected according to the loop-shaping technique. \\
Open-loop frequency analysis of the two systems in Fig~\ref{fig:openloop}. also reveals some helpful information. In addition to revealing the higher-order harmonic contents of the output of the CLOC system, it can be seen that at 12 Hz, where $\psi$ crosses zero line, higher-order harmonics show a notch-like behaviour, and their magnitude will be zero. It also shows that the positive gain of $\psi$ created a positive phase slope for CgLp. It should be noted that the discrepancy between the slope of $\psi$ and the phase of CLOC at the beginning and the end of $[\omega_l,\omega_h]$, is because $[\omega_d,\omega_t]$ is smaller than $[\omega_l,\omega_h]$. \\
The step response of the CLOC and PID for bandwidth of 50~Hz, 100~Hz and 200~Hz has been obtained using the Simulink environment of Matlab and depicted in Fig.~\ref{fig:step}. It should be noted that bandwidth change has been created only through changing $k_p$, and no other parameter has been changed. This figure reveals that CLOC shows a lower overshoot in all cases than PID even when two controllers have the same phase margin. It can be explained through the reset nature of the CLOC. As it is expected from Fig.~\ref{fig:openloop}, when bandwidth changes from 100~Hz to 200~Hz, the phase margin of PID will be reduced and thus, its overshoot increases, while the phase margin of CLOC increases and thus, its overshoot reduces. When bandwidth is changed to 50~Hz, both controllers lose phase margin and show an increase in overshoot; however, CLOC still shows a lower overshoot than PID even when its phase margin is lower than PID.\\
The sensitivity plot of the control system can be used to validate the improvement in steady-state performance. However, because reset control systems are nonlinear, therefore, sensitivity plots for them must be estimated; the sensitivity plot obtained using the DF approximation may not be accurate~\cite{saikumar2021loop}.  In order to compute the sensitivity plot more precisely, a series of simulations for tracking sinusoidal waves with different frequencies were performed, and the $\frac{\Vert e(t) \Vert 2}{\Vert r(t) \Vert 2}$ was shown for both controllers in Fig.~\ref{fig:sensitivity}.\\
As it was expected, because of the higher open-loop gain at lower frequencies (see Fig.~\ref{fig:openloop}), CLOC shows a lower tracking error compared to PID and also a lower peak of sensitivity. At last, to see the steady-state time-domain results, the sinusoidal tracking performance of the controllers for $r(t)=\sin(2\pi t)$ is depicted in Fig.~\ref{fig:sin}. 
\section{CONCLUSIONS}

This paper introduced a reset controller based on the structure of a CgLp with a shaped reset signal, which could approximate a complex-order transfer function. For this purpose, a shaping filter for the reset signal was designed, which could alter the DF of the reset controller to achieve the negative gain slope along with a positive phase slope. Furthermore, a tuning method was introduced to tune the gain slope and, more importantly, phase slope. In order to illustrate the possible improvements of using such a controller, a comparison between CLOC and PID was made in frequency and time domain. It was shown that CLOC could achieve the same phase margin as PID with a weaker linear tamed differentiator and showed a lower overshoot even with the same phase margin as PID. The weaker linear tamed differentiator also facilitated the CLOC to show higher gain in lower frequencies, thus showing a lower sinusoidal tracking error.\\
Implementation of the proposed controller in practice in the presence of noise and disturbance for a more general plant can be a follow-up for this study. 

\addtolength{\textheight}{-8cm}   





\bibliographystyle{IEEEtran}

\bibliography{ref}

\begin{thebibliography}{10}
\providecommand{\url}[1]{#1}
\csname url@samestyle\endcsname
\providecommand{\newblock}{\relax}
\providecommand{\bibinfo}[2]{#2}
\providecommand{\BIBentrySTDinterwordspacing}{\spaceskip=0pt\relax}
\providecommand{\BIBentryALTinterwordstretchfactor}{4}
\providecommand{\BIBentryALTinterwordspacing}{\spaceskip=\fontdimen2\font plus
\BIBentryALTinterwordstretchfactor\fontdimen3\font minus
  \fontdimen4\font\relax}
\providecommand{\BIBforeignlanguage}[2]{{%
\expandafter\ifx\csname l@#1\endcsname\relax
\typeout{** WARNING: IEEEtran.bst: No hyphenation pattern has been}%
\typeout{** loaded for the language `#1'. Using the pattern for}%
\typeout{** the default language instead.}%
\else
\language=\csname l@#1\endcsname
\fi
#2}}
\providecommand{\BIBdecl}{\relax}
\BIBdecl

\bibitem{aastrom2000limitations}
K.~J. {\AA}str{\"o}m, ``Limitations on control system performance,''
  \emph{European Journal of Control}, vol.~6, no.~1, pp. 2--20, 2000.

\bibitem{schmidt2020design}
R.~M. Schmidt, G.~Schitter, and A.~Rankers, \emph{The design of high
  performance mechatronics: {High-Tech} functionality by multidisciplinary
  system integration}.\hskip 1em plus 0.5em minus 0.4em\relax {IOS} Press,
  2020.

\bibitem{freudenberg2000survey}
J.~Freudenberg, R.~Middleton, and A.~Stefanpoulou, ``A survey of inherent
  design limitations,'' in \emph{Proceedings of the 2000 American Control
  Conference. ACC (IEEE Cat. No. 00CH36334)}, vol.~5.\hskip 1em plus 0.5em
  minus 0.4em\relax IEEE, 2000, pp. 2987--3001.

\bibitem{oustaloup2000frequency}
A.~Oustaloup, F.~Levron, B.~Mathieu, and F.~M. Nanot, ``Frequency-band complex
  noninteger differentiator: characterization and synthesis,'' \emph{IEEE
  Transactions on Circuits and Systems I: Fundamental Theory and Applications},
  vol.~47, no.~1, pp. 25--39, 2000.

\bibitem{sabatier2015fractional}
J.~Sabatier, P.~Lanusse, P.~Melchior, and A.~Oustaloup, ``Fractional order
  differentiation and robust control design,'' \emph{Intelligent systems,
  control and automation: science and engineering}, vol.~77, pp. 13--18, 2015.

\bibitem{oustaloup1991commande}
A.~Oustaloup, \emph{La commande CRONE: commande robuste d'ordre non
  entier}.\hskip 1em plus 0.5em minus 0.4em\relax Hermes, 1991.

\bibitem{moghadam2018tuning}
M.~G. Moghadam, F.~Padula, and L.~Ntogramatzidis, ``Tuning and performance
  assessment of complex fractional-order pi controllers,''
  \emph{IFAC-PapersOnLine}, vol.~51, no.~4, pp. 757--762, 2018.

\bibitem{clegg1958nonlinear}
J.~Clegg, ``A nonlinear integrator for servomechanisms,'' \emph{Transactions of
  the American Institute of Electrical Engineers, Part II: Applications and
  Industry}, vol.~77, no.~1, pp. 41--42, 1958.

\bibitem{valerio2019reset}
D.~Val{\'e}rio, N.~Saikumar, A.~A. Dastjerdi, N.~Karbasizadeh, and S.~H.
  HosseinNia, ``Reset control approximates complex order transfer functions,''
  \emph{Nonlinear Dynamics}, vol.~97, no.~4, pp. 2323--2337, 2019.

\bibitem{saikumar2019complex}
N.~Saikumar, D.~Val{\'e}rio, and S.~H. HosseinNia, ``Complex order control for
  improved loop-shaping in precision positioning,'' in \emph{2019 IEEE 58th
  Conference on Decision and Control (CDC)}.\hskip 1em plus 0.5em minus
  0.4em\relax IEEE, 2019, pp. 7956--7962.

\bibitem{banos2011reset}
A.~Ba{\~n}os and A.~Barreiro, \emph{Reset control systems}.\hskip 1em plus
  0.5em minus 0.4em\relax Springer Science \& Business Media, 2011.

\bibitem{horowitz1975non}
I.~Horowitz and P.~Rosenbaum, ``Non-linear design for cost of feedback
  reduction in systems with large parameter uncertainty,'' \emph{International
  Journal of Control}, vol.~21, no.~6, pp. 977--1001, 1975.

\bibitem{krishnan1974synthesis}
K.~Krishnan and I.~Horowitz, ``Synthesis of a non-linear feedback system with
  significant plant-ignorance for prescribed system tolerances,''
  \emph{International Journal of Control}, vol.~19, no.~4, pp. 689--706, 1974.

\bibitem{hazeleger2016second}
L.~Hazeleger, M.~Heertjes, and H.~Nijmeijer, ``Second-order reset elements for
  stage control design,'' in \emph{2016 American Control Conference
  (ACC)}.\hskip 1em plus 0.5em minus 0.4em\relax IEEE, 2016, pp. 2643--2648.

\bibitem{banos2007definition}
A.~Ba{\~n}os and A.~Vidal, ``Definition and tuning of a {PI}+ {CI} reset
  controller,'' in \emph{2007 European Control Conference (ECC)}.\hskip 1em
  plus 0.5em minus 0.4em\relax IEEE, 2007, pp. 4792--4798.

\bibitem{beker2004fundamental}
O.~Beker, C.~Hollot, Y.~Chait, and H.~Han, ``Fundamental properties of reset
  control systems,'' \emph{Automatica}, vol.~40, no.~6, pp. 905--915, 2004.

\bibitem{saikumar2019constant}
N.~{Saikumar}, R.~K. {Sinha}, and S.~H. {HosseinNia}, ``{“Constant in Gain
  Lead in Phase”} element– application in precision motion control,''
  \emph{IEEE/ASME Transactions on Mechatronics}, vol.~24, no.~3, pp.
  1176--1185, 2019.

\bibitem{karbasizadeh2021fractional}
N.~Karbasizadeh, N.~Saikumar, and S.~H. {HosseinNia},
  ``\BIBforeignlanguage{English}{Fractional-order single state reset
  element},'' \emph{\BIBforeignlanguage{English}{Nonlinear Dynamics}}, vol.
  104, no.~1, pp. 413--427, 2021.

\bibitem{karbasizadeh2020benefiting}
N.~Karbasizadeh, A.~A. Dastjerdi, N.~Saikumar, D.~Valério, and S.~H.
  HosseinNia, ``Benefiting from linear behaviour of a nonlinear reset-based
  element at certain frequencies,'' in \emph{2020 Australian and New Zealand
  Control Conference (ANZCC)}, 2020, pp. 226--231.

\bibitem{dastjerdi2021frequency}
A.~A. Dastjerdi and S.~H. {HosseinNia}, ``A frequency-domain tuning method for
  a class of reset control systems,'' \emph{IEEE Access}, vol.~9, pp.
  40\,950--40\,962, 2021.

\bibitem{guo2009frequency}
Y.~Guo, Y.~Wang, and L.~Xie, ``Frequency-domain properties of reset systems
  with application in hard-disk-drive systems,'' \emph{IEEE Transactions on
  Control Systems Technology}, vol.~17, no.~6, pp. 1446--1453, 2009.

\bibitem{nuij2006higher}
P.~Nuij, O.~Bosgra, and M.~Steinbuch, ``Higher-order sinusoidal input
  describing functions for the analysis of non-linear systems with harmonic
  responses,'' \emph{Mechanical Systems and Signal Processing}, vol.~20, no.~8,
  pp. 1883--1904, 2006.

\bibitem{saikumar2021loop}
N.~Saikumar, K.~Heinen, and S.~H. HosseinNia, ``Loop-shaping for reset control
  systems: A higher-order sinusoidal-input describing functions approach,''
  \emph{Control Engineering Practice}, vol. 111, p. 104808, 2021.

\bibitem{dastjerdi2020closed}
\BIBentryALTinterwordspacing
A.~A. Dastjerdi, N.~Saikumar, D.~Val{\'e}rio, and S.~H. HosseinNia,
  ``Closed-loop frequency analyses of reset systems,'' \emph{arXiv preprint
  arXiv:2001.10487}, 2020. [Online]. Available:
  \url{https://arxiv.org/abs/2001.10487}
\BIBentrySTDinterwordspacing

\bibitem{Guo:2015}
Y.~Guo, L.~Xie, and Y.~Wang, \emph{Analysis and Design of Reset Control
  Systems}.\hskip 1em plus 0.5em minus 0.4em\relax Institution of Engineering
  and Technology, 2015.

\bibitem{nevsic2008stability}
D.~Ne{\v{s}}i{\'c}, L.~Zaccarian, and A.~R. Teel, ``Stability properties of
  reset systems,'' \emph{Automatica}, vol.~44, no.~8, pp. 2019--2026, 2008.

\bibitem{dastjerdi2021newstability}
A.~A. Dastjerdi, A.~Astolfi, and S.~H. HosseinNia, ``Frequency domain stability
  method for reset systems,'' 2021.

\bibitem{podlubny1998fractional}
I.~Podlubny, \emph{Fractional differential equations: an introduction to
  fractional derivatives, fractional differential equations, to methods of
  their solution and some of their applications}.\hskip 1em plus 0.5em minus
  0.4em\relax Elsevier, 1998.

\bibitem{samko1993fractional}
S.~G. Samko, A.~A. Kilbas, and O.~I. Marichev, ``Fractional integrals and
  derivatives, translated from the 1987 russian original,'' 1993.

\bibitem{valerio2013introduction}
D.~Val{\'e}rio and J.~S. Da~Costa, \emph{An introduction to fractional
  control}.\hskip 1em plus 0.5em minus 0.4em\relax IET, 2013, vol.~91.

\bibitem{karbasizadeh2020bandpassing}
\BIBentryALTinterwordspacing
N.~Karbasizadeh, A.~A. Dastjerdi, N.~Saikumar, and S.~H. HosseinNia,
  ``Band-passing nonlinearity in reset elements,'' \emph{IEEE Transactions on
  Control Systems Technology}, In press. [Online]. Available:
  \url{https://arxiv.org/abs/2009.06091}
\BIBentrySTDinterwordspacing

\bibitem{bahnamiri2020tuning}
\BIBentryALTinterwordspacing
M.~S. Bahnamiri, N.~Karbasizadeh, A.~A. Dastjerdi, N.~Saikumar, and S.~H.
  HosseinNia, ``Tuning of {CgLp} based reset controllers: Application in
  precision positioning systems,'' \emph{{IFAC}-{PapersOnLine}}, vol.~53,
  no.~2, pp. 8997--9004, 2020. [Online]. Available:
  \url{https://doi.org/10.1016/j.ifacol.2020.12.2017}
\BIBentrySTDinterwordspacing

\bibitem{valerio2013variable}
D.~Val{\'e}rio and J.~S{\'a}~da Costa, ``Variable order fractional
  controllers,'' \emph{Asian Journal of Control}, vol.~15, no.~3, pp. 648--657,
  2013.

\end{thebibliography}

\end{document}